

\def\oneandahalfspace{\baselineskip=\normalbaselineskip
  \multiply\baselineskip by 3 \divide\baselineskip by 2}

\parskip=\medskipamount
\overfullrule=0pt
\raggedbottom
\def\normalparindent{24pt}
\nopagenumbers
\footline={\ifnum\pageno=1{\hfil}\else{\hfil\rm\folio\hfil}\fi}
\def\endpage{\vfill\eject}
\def\beginlinemode{\endmode\begingroup\parskip=0pt
                   \obeylines\def\\{\par}\def\endmode{\par\endgroup}}
\def\beginparmode{\endmode\begingroup \def\endmode{\par\endgroup}}
\let\endmode=\par
\def\raggedcenter{
                  \leftskip=2em plus 6em \rightskip=\leftskip
                  \parindent=0pt \parfillskip=0pt \spaceskip=.3333em
                  \xspaceskip=.5em\pretolerance=9999 \tolerance=9999
                  \hyphenpenalty=9999 \exhyphenpenalty=9999 }
\def\\{\cr}
\let\rawfootnote=\footnote\def\footnote#1#2{{\parindent=0pt\parskip=0pt
        \rawfootnote{#1}{#2\hfill\vrule height 0pt depth 6pt width 0pt}}}
\def\title{\null\vskip 3pt plus 0.2fill\beginlinemode\raggedcenter\bf}
\def\author{\vskip 3pt plus 0.2fill \beginlinemode\raggedcenter}
\def\affil{\vskip 3pt plus 0.1fill\beginlinemode\raggedcenter\it}
\def\abstract{\vskip 3pt plus 0.3fill \beginparmode{\noindent  ABSTRACT:~}  }
\def\endtitlepage{\endpage\body}
\def\body{\beginparmode\parindent=\normalparindent}
\def\head#1{\par\goodbreak{\immediate\write16{#1}
           {\noindent\bf #1}\par}\nobreak\nobreak}
\def\subhead#1{\par\goodbreak{\immediate\write16{#1}
           {\noindent\bf #1}\par}\nobreak\nobreak}
\def\refto#1{$^{[#1]}$}
\def\ref#1{Ref.~#1}
\def\Ref#1{Ref.~#1}\def\cite#1{{#1}}\def\[#1]{[\cite{#1}]}

\def\(#1){(\call{#1})}
\def\call#1{{#1}}\def\taghead#1{{#1}}
\def\references{\head{REFERENCES}\beginparmode\frenchspacing\parskip=0pt}
\gdef\refis#1{\item{#1.\ }}
\gdef\journal#1,#2,#3,#4.{{\sl #1~}{\bf #2} (#4) #3}
\def\endreferences{\body}
\def\endit{\endmode\vfill\supereject}\let\endpaper=\endit

\def\prd{\journal Phys. Rev. D,}
\def\prl{\journal Phys. Rev. Lett.,}

\def\npb{\journal Nucl. Phys. B,}

\def\plb{\journal Phys. Lett. B,}

\def\cqg{\journal Class. Quantum Grav.,}

\def\gsim{\mathrel{\raise.3ex\hbox{$>$\kern-.75em\lower1ex\hbox{$\sim$}}}}
\def\lsim{\mathrel{\raise.3ex\hbox{$<$\kern-.75em\lower1ex\hbox{$\sim$}}}}
\def\square{\kern1pt\vbox{\hrule height 0.6pt\hbox{\vrule width 0.6pt\hskip 3pt
   \vbox{\vskip 6pt}\hskip 3pt\vrule width 0.6pt}\hrule height 0.6pt}\kern1pt}
\def\sla{\raise.15ex\hbox{$/$}\kern-.72em}

\catcode`@=11
\newcount\r@fcount \r@fcount=0\newcount\r@fcurr
\immediate\newwrite\reffile\newif\ifr@ffile\r@ffilefalse
\def\w@rnwrite#1{\ifr@ffile\immediate\write\reffile{#1}\fi\message{#1}}
\def\writer@f#1>>{}
\def\referencefile{\r@ffiletrue\immediate\openout\reffile=\jobname.ref%
  \def\writer@f##1>>{\ifr@ffile\immediate\write\reffile%
    {\noexpand\refis{##1} = \csname r@fnum##1\endcsname = %
     \expandafter\expandafter\expandafter\strip@t\expandafter%
     \meaning\csname r@ftext\csname r@fnum##1\endcsname\endcsname}\fi}%
  \def\strip@t##1>>{}}

\def\citeall#1{\xdef#1##1{#1{\noexpand\cite{##1}}}}
\def\cite#1{\each@rg\citer@nge{#1}}
\def\each@rg#1#2{{\let\thecsname=#1\expandafter\first@rg#2,\end,}}
\def\first@rg#1,{\thecsname{#1}\apply@rg}
\def\apply@rg#1,{\ifx\end#1\let\next=\relax%
\else,\thecsname{#1}\let\next=\apply@rg\fi\next}%
\def\citer@nge#1{\citedor@nge#1-\end-}
\def\citer@ngeat#1\end-{#1}
\def\citedor@nge#1-#2-{\ifx\end#2\r@featspace#1
  \else\citel@@p{#1}{#2}\citer@ngeat\fi}
\def\citel@@p#1#2{\ifnum#1>#2{\errmessage{Reference range #1-#2\space is bad.}
    \errhelp{If you cite a series of references by the notation M-N, then M and
    N must be integers, and N must be greater than or equal to M.}}\else%
{\count0=#1\count1=#2\advance\count1 by1\relax\expandafter\r@fcite\the\count0,%
  \loop\advance\count0 by1\relax
    \ifnum\count0<\count1,\expandafter\r@fcite\the\count0,%
  \repeat}\fi}
\def\r@featspace#1#2 {\r@fcite#1#2,}    \def\r@fcite#1,{\ifuncit@d{#1}
    \expandafter\gdef\csname r@ftext\number\r@fcount\endcsname%
    {\message{Reference #1 to be supplied.}\writer@f#1>>#1 to be supplied.\par
     }\fi\csname r@fnum#1\endcsname}
\def\ifuncit@d#1{\expandafter\ifx\csname r@fnum#1\endcsname\relax%
\global\advance\r@fcount by1%
\expandafter\xdef\csname r@fnum#1\endcsname{\number\r@fcount}}
\let\r@fis=\refis   \def\refis#1#2#3\par{\ifuncit@d{#1}%
    \w@rnwrite{Reference #1=\number\r@fcount\space is not cited up to now.}\fi%
  \expandafter\gdef\csname r@ftext\csname r@fnum#1\endcsname\endcsname%
  {\writer@f#1>>#2#3\par}}
\def\r@ferr{\endreferences\errmessage{I was expecting to see
\noexpand\endreferences before now;  I have inserted it here.}}
\let\r@ferences=\references
\def\references{\r@ferences\def\endmode{\r@ferr\par\endgroup}}
\let\endr@ferences=\endreferences
\def\endreferences{\r@fcurr=0{\loop\ifnum\r@fcurr<\r@fcount
    \advance\r@fcurr by 1\relax\expandafter\r@fis\expandafter{\number\r@fcurr}%
    \csname r@ftext\number\r@fcurr\endcsname%
  \repeat}\gdef\r@ferr{}\endr@ferences}
\let\r@fend=\endpaper\gdef\endpaper{\ifr@ffile
\immediate\write16{Cross References written on []\jobname.REF.}\fi\r@fend}
\catcode`@=12
\citeall\refto\citeall\ref\citeall\Ref
\catcode`@=11
\newcount\tagnumber\tagnumber=0
\immediate\newwrite\eqnfile\newif\if@qnfile\@qnfilefalse
\def\write@qn#1{}\def\writenew@qn#1{}
\def\w@rnwrite#1{\write@qn{#1}\message{#1}}
\def\@rrwrite#1{\write@qn{#1}\errmessage{#1}}
\def\taghead#1{\gdef\t@ghead{#1}\global\tagnumber=0}
\def\t@ghead{}\expandafter\def\csname @qnnum-3\endcsname
  {{\t@ghead\advance\tagnumber by -3\relax\number\tagnumber}}
\expandafter\def\csname @qnnum-2\endcsname
  {{\t@ghead\advance\tagnumber by -2\relax\number\tagnumber}}
\expandafter\def\csname @qnnum-1\endcsname
  {{\t@ghead\advance\tagnumber by -1\relax\number\tagnumber}}
\expandafter\def\csname @qnnum0\endcsname
  {\t@ghead\number\tagnumber}
\expandafter\def\csname @qnnum+1\endcsname
  {{\t@ghead\advance\tagnumber by 1\relax\number\tagnumber}}
\expandafter\def\csname @qnnum+2\endcsname
  {{\t@ghead\advance\tagnumber by 2\relax\number\tagnumber}}
\expandafter\def\csname @qnnum+3\endcsname
  {{\t@ghead\advance\tagnumber by 3\relax\number\tagnumber}}
\def\equationfile{\@qnfiletrue\immediate\openout\eqnfile=\jobname.eqn%
  \def\write@qn##1{\if@qnfile\immediate\write\eqnfile{##1}\fi}
  \def\writenew@qn##1{\if@qnfile\immediate\write\eqnfile
    {\noexpand\tag{##1} = (\t@ghead\number\tagnumber)}\fi}}
\def\callall#1{\xdef#1##1{#1{\noexpand\call{##1}}}}
\def\call#1{\each@rg\callr@nge{#1}}
\def\each@rg#1#2{{\let\thecsname=#1\expandafter\first@rg#2,\end,}}
\def\first@rg#1,{\thecsname{#1}\apply@rg}
\def\apply@rg#1,{\ifx\end#1\let\next=\relax%
\else,\thecsname{#1}\let\next=\apply@rg\fi\next}
\def\callr@nge#1{\calldor@nge#1-\end-}\def\callr@ngeat#1\end-{#1}
\def\calldor@nge#1-#2-{\ifx\end#2\@qneatspace#1 %
  \else\calll@@p{#1}{#2}\callr@ngeat\fi}
\def\calll@@p#1#2{\ifnum#1>#2{\@rrwrite{Equation range #1-#2\space is bad.}
\errhelp{If you call a series of equations by the notation M-N, then M and
N must be integers, and N must be greater than or equal to M.}}\else%
{\count0=#1\count1=#2\advance\count1 by1\relax\expandafter\@qncall\the\count0,%
  \loop\advance\count0 by1\relax%
    \ifnum\count0<\count1,\expandafter\@qncall\the\count0,  \repeat}\fi}
\def\@qneatspace#1#2 {\@qncall#1#2,}
\def\@qncall#1,{\ifunc@lled{#1}{\def\next{#1}\ifx\next\empty\else
  \w@rnwrite{Equation number \noexpand\(>>#1<<) has not been defined yet.}
  >>#1<<\fi}\else\csname @qnnum#1\endcsname\fi}
\let\eqnono=\eqno\def\eqno(#1){\tag#1}\def\tag#1$${\eqnono(\displayt@g#1 )$$}
\def\aligntag#1\endaligntag  $${\gdef\tag##1\\{&(##1 )\cr}\eqalignno{#1\\}$$
  \gdef\tag##1$${\eqnono(\displayt@g##1 )$$}}
\def\eqalignno#1{\displ@y \tabskip\centering
  \halign to\displaywidth{\hfil$\displaystyle{##}$\tabskip\z@skip
    &$\displaystyle{{}##}$\hfil\tabskip\centering
    &\llap{$\displayt@gpar##$}\tabskip\z@skip\crcr
    #1\crcr}}
\def\displayt@gpar(#1){(\displayt@g#1 )}
\def\displayt@g#1 {\rm\ifunc@lled{#1}\global\advance\tagnumber by1
        {\def\next{#1}\ifx\next\empty\else\expandafter
        \xdef\csname @qnnum#1\endcsname{\t@ghead\number\tagnumber}\fi}%
  \writenew@qn{#1}\t@ghead\number\tagnumber\else
        {\edef\next{\t@ghead\number\tagnumber}%
        \expandafter\ifx\csname @qnnum#1\endcsname\next\else
        \w@rnwrite{Equation \noexpand\tag{#1} is a duplicate number.}\fi}%
  \csname @qnnum#1\endcsname\fi}
\def\ifunc@lled#1{\expandafter\ifx\csname @qnnum#1\endcsname\relax}
\let\@qnend=\end\gdef\end{\if@qnfile
\immediate\write16{Equation numbers written on []\jobname.EQN.}\fi\@qnend}
\catcode`@=12


\magnification=1200
\parskip=\smallskipamount

\hfill Preprint KONS-RGKU-94-06
\title
Bremsstrahlung in the gravitational field of a cosmic string
\author
J\"urgen Audretsch$^1$, Ulf Jasper$^{1,}$\footnote{$^\dagger$}{{\it e-mail:
ulf@spock.physik.uni-konstanz.de}}, and Vladimir D.
Skarzhinsky$^{2,1,}$\footnote{$^\ddagger$}{{\it e-mail:
skarzh@npad.fian.msk.su}}

\affil $^1$Fakult\"at f\"ur Physik der Universit\"at Konstanz
Postfach 5560 M 674, D 78434 Konstanz, Germany

\affil $^2$P. N. Lebedev Physical Institute
Leninsky prospect 53, Moscow 117924, Russia

\abstract

In the framework of QED we investigate the bremsstrahlung process
for an electron passing by a straight static cosmic string. This
process is precluded in empty Minkowski space-time by energy and
momentum conservation laws. It happens in the presence of the cosmic
string as a consequence of the conical structure of space, in spite
of the flatness of the metric. The cross section and emitted electromagnetic
energy are computed and analytic expressions are found for different energies
of the incoming electron. The energy interval is divided in three parts
depending on whether the energy is just above electron rest mass $M$,
much larger than $M$, or exceeds $M/\delta$, with $\delta$ the string
mass per unit length in Planck units. We compare our results with those
of scalar QED and classical electrodynamics and also with conic pair production
process computed earlier.

\bigskip
\noindent PACS numbers:  04.62.+v, 03.70.+k, 98.80.Cq
\vfill
\noindent To appear in Phys. Rev. D.
\endtitlepage\vfil\eject\body

\head{1. Introduction}
\vskip0.5cm

The bremsstrahlung process is a rather common process well known since
earlier times of QED \refto{Bethe34}. It happens usually when a charged
particle changes its momentum in collision with obstacles such as other
particles or due to an acceleration in electromagnetic fields. This quantum
process has its transparent classical counterpart permitting to compare
classical and QED calculations. Charged particles moving freely in flat
space-time do not radiate.

A quite different situation occurs in curved space-time. As was shown in
\refto{DeWitt60}, a charged particle moving on geodesics does radiate. The
radiation occurs not so much because of the non zero curvature, but rather due
to the falldown of the Huygens principle in curved space-time. It was
manifested recently in \refto{Frolov88} by direct calculation of the classical
bremsstrahlung radiation of an electric charge in the gravitational field of a
straight static cosmic string and also for scalar and gravitational radiation
\refto{Aliev89a,Aliev89b}. These authors have given analytic expressions for
the total radiation energy emitted by classical point charge moving on
geodesics of cosmic string space-time. The quantum bremsstrahlung process for a
scalar field model was treated in details in \refto{Audretsch91a,Audretsch91b}.
The aim of this paper is to evaluate the total cross section as well as the
radiation energy emitted by an electron passing by the cosmic string and to
compare our results with those found early for classical and scalar quantum
electrodynamics.

The space-time of a straight static cosmic string is locally flat except for
the string itself where the Riemann tensor is concentrated. The metric around
the string that lies along the z-axis reads, in cylindrical coordinates
\refto{Vilenkin85,Gott85}:
$$
ds^2=dt^2-d\rho^2-\rho^2d\theta^2-dz^2\;.\eqno(m)$$
The metric is the same as in Minkowski space, but here the periodicity
of the angular coordinate is within the range
$$
0\leq\theta\leq {2\pi\over\nu},\quad{\rm with}\quad\nu=(1-4G\mu)^{-1}.
\eqno(d)$$
$\mu$ is the mass per unit length of string, and $G$ is Newton's constant. The
space-like sections around the string have the topology of a cone with the
vertex at the core and with deficit angle $8\pi G\mu$. This dimensionless
quantity measures the strength of the gravitational effects of the string.

Without any Newtonian gravitational field around, the string can produce
physical effects in a hidden way, via the topological structure of the
space-time as well as massive particles do this in 2+1 dimensional gravity
\refto{Henneaux84,Deser84,Deser88}.
It bears some resemblance with the Aharonov-Bohm effect \refto{Aharonov59} at
least in its topological aspects. The close analogy between these effects,
though they are of a different nature, was noticed by many authors
\refto{Ford81,Bezzera87,Alford89a,Alford89b,Gerbert89a,Perkins91a}. Recently
the Aharonov-Bohm interaction of fermions with the pure gauge potentials around
cosmic strings has been shown to lead to their significantly large interaction
with matter fields concentrated in the cosmic string core
\refto{Alford89a,Alford89b,Gerbert89a,Perkins91a}. We notice that it does not
happen due to gravitational interaction alone.

Classically, the gravitational effects of a straight cosmic string become
manifest when two or more particles move along opposite sides of a string.
Initially parallel trajectories converge as they move past the string, which
acts as a gravitational lens \refto{Vilenkin81,Vilenkin85}. When moving
thorough surrounded matter, a string produces wakes besides itself
\refto{Vilenkin85}, and its relative motion introduces a steplike discontinuity
in the observed microwave background temperature \refto{Gott85,Kaiser84}. Other
classical effects may affect a single particle: the conical structure of the
space-time induces a self-force, both gravitational as well as electrostatic,
on a test particle around a string \refto{Linet86,Smith90}. Being dependent on
the distance from the string it is the effect which leads to the classical
bremsstrahlung radiation from the particle passing by the string
\refto{Frolov88, Aliev89a}. Also, a charged particle radiates when it suffers
Aharonov-Bohm scattering \refto{Serebryany89}. Among its quantum effects, a
string polarizes the vacuum that surrounds it, in a way analogous to the
Casimir effect between two plane conductors forming a wedge
\refto{Helliwel86,Frolov87,Dowker87,Hiscock87}.

The conical structure of the cosmic string space-time prevents the invariance
under translations in the plane perpendicular to the string and is the source
of non-conservation for transverse momentum. It allows for another type of
effects on quantum fields. Firstly it was noticed in \refto{Harari90} for pair
production in the context of a simplified model based on a scalar field theory.
This process was analyzed in more details in \refto{Audretsch91c,Harari91}. In
the previous paper \refto{Skarzh93} the cross section for pair production by a
single photon in the cosmic string background was evaluated  and some of its
potential consequences were discussed in the framework of QED. An heuristic,
semiclassical explanation for the mechanism underlying different quantum
effects with free particles around the string follows. The vacuum is full of
virtual particles which are continuously created and quickly disappear. A free
particle is not able to make them real in empty Minkowski space-time, even if
it has sufficient energy, since otherwise momentum conservation law would be
violated. The presence of the string allows to take off the momentum excess and
make them real if outgoing particles move along opposite sides of the string.
This picture proves to be helpful to understand qualitatively some of the
quantitative results for processes in study.

The paper is organized as follows: In section 2 we review briefly the Dirac and
Maxwell equations in the cosmic string space-time with reference for details to
the previous paper \refto{Skarzh93}. In section 3 we evaluate the first order
matrix element for the bremsstrahlung process in the cosmic string
gravitational background. The analytic expressions for differential probability
and partial cross section are found and analyzed. The radiated energy emitted
by a free electron is computed in section 4 where it is also compared with the
classical one. We present analytic approximations valid at different energy
regimes. We conclude in section 5 with a discussion about the obtained results.

\vskip1cm

\head{2. Dirac and Maxwell fields in the cosmic string space-time}
\vskip0.5cm

The Dirac and Maxwell fields in the space-time of a straight cosmic string with
the metric \(m) have been treated in a previous paper \refto{Skarzh93}. In this
section we review briefly their main properties and present in more details the
physical states which describe transverse, physical photons.

The electron-positron field operator $\psi(x,t)$ obeys the free Dirac equation
written in cylindrical coordinates for the cosmic string metric \(m) as
$$
\left( i\left[\gamma^{0}\partial_{t} + \gamma^{3}\partial_{z} +
\gamma^{\rho}(\partial_{\rho} - {{\nu -1} \over {2\rho}}) +
\gamma^{\theta}\partial_{\theta}\right] - M \right) \psi = 0\;.
\eqno(dirac)$$
It differs from the usual one by the spin connection term $\gamma^{\rho}{\nu-1
\over {2\rho}}$. The equation \(dirac) was investigated early in 2+1
dimensions, with two-component spinors, both in a conical space
\refto{Gerbert89b} as well as in the Aharonov-Bohm external field
\refto{Alford89a,Alford89b,Gerbert89b,Perkins91a,Hagen90,Voropaev91,Kogan91}.
To specify the solution to the Dirac equation \(dirac) we introduce a complete
set of commuting operators
$$\eqalign{
\hat{H}\psi =& E\psi,\quad \hat{p}_z\psi = p_3\psi,\cr
\hat{J}_3\psi =& (-i\partial_{\theta}+{\nu \over 2} \Sigma_3)\psi = j_3\psi, \;
j_3=\nu j, \; j=l+{1\over 2},\cr
S_t\psi=& \;s\psi,\; \hat{S}_t = {1\over \sqrt{E^2-M^2}}\Sigma_ip_i. \cr}$$
and use their eigenvalues to label the quantum states of electrons. Here
$\hat{H}, \hat{p}_3, \hat{J}_3 \;{\rm and}\; \hat{S}_3$ correspond to energy,
z-momentum, z-projection of total angular momentum and helicity operators,
respectively.

We need for the following calculations the electron wave functions only. They
are presented by the cylindrical modes:
$$
\psi_{e}(j,x) = {\sqrt{\nu} \over 2\pi}N_{e}\exp(-iEt + ip_{3} z)
\exp\left(i{\pi \over 2}|l|\right) \cdot \pmatrix{ u\cr
v\cr}\;\eqno(es) $$
with the two-component spinors $u,v$ given by
$$
u = {1 \over \sqrt{E - M}}\cdot
\pmatrix{ J_{\alpha_-}(p_{\perp}\rho)\exp(i\nu l\theta)\cr
{i\epsilon_{l}sp_{\perp} \over p +
sp_3}J_{\alpha_+}(p_{\perp}\rho)\exp(i\nu(l +1)\theta)\cr}$$
$$
v = {1 \over \sqrt{E + M}}\cdot
\pmatrix{ sJ_{\alpha_-}(p_{\perp}\rho)\exp(i\nu l\theta)\cr
{i\epsilon_{l}p_{\perp} \over p +
sp_3}J_{\alpha_+}(p_{\perp}\rho)\exp(i\nu(l +1)\theta)\cr}$$
and the normalization constant $N_{e}={\sqrt{p(p+sp_3)}\over{2\sqrt{E}}}.$ The
index of the Bessel function is $\alpha_{\mp} = |\nu j \mp{1/2}|$ and its
argument depends on the transverse momentum $p_{\perp}=\sqrt{p^2-p^2_3}
=\sqrt{E^2-M^2-p^2_3}$. $\epsilon_l ={\rm sign}(l)$ and $s=\pm 1.$ We added a
phase factor to \(es) to simplify the calculations.

The normalization condition for these normal modes is
$$
\int dx
\psi^{\dagger}_{e}(j,x) \psi_{e}(j^{\prime},x) = \delta_{j, j^{
\prime}}={\delta_{s, s^{\prime}}} {\delta_{l,  l^{ \prime}}}
{ \delta(p_{3} - p_3^\prime)} {\delta(p_{ \perp} -  p_{ \perp}^{ \prime}) \over
\sqrt{p_{\perp} p_{ \perp}^{ \prime}}}\;.\eqno(dn) $$
We denote collectively by $j$ or $j^\prime$ the quantum numbers of a given
state, and integration is also collectively denoted as
$$
\int d \mu_{j}
= { \sum_{l= - \infty}^{ \infty}}{ \int_{- \infty}^{  \infty}
{dp_{3}}}{ \int_{0}^{ \infty} p_{ \perp} dp_{ \perp}}\;. $$

In order to make a comparison with classical and scalar calculations possible
we will discuss these modes now. This can be done in analogy with the case of
scalar quantum fields as it was developed in \refto{Audretsch91a}. The modes
\(es) give a particle density at $p_3=0$,
$$
j^0(\rho)={\nu\over 8\pi^2}[J^2_{\alpha_-}(p_{\perp}\rho)+
J^2_{\alpha_+}(p_{\perp}\rho)],$$
which behaves for $j\nu \ll p_{\perp}\rho$ like $j^0(\rho)\approx\nu /(4
\pi^3p{\perp}\rho)$ and for $j\nu \gg p_{\perp}\rho$ like $(\nu /8\pi^2)\cdot
(p_{\perp}\rho /2)^{2j}\Gamma^{-2}(j+1).$
The transition between these regimes takes place around $\rho_{\rm min}=\nu
j/p_{\perp}$ which is the radius of closest approach of a classical test
particle of radial momentum $p_{\perp}$ and $z$-angular momentum $j$. For
$\rho_{\rm min}<\rho$ the particle density is very small so that one may speak
of a {\it localized absence} \refto{Audretsch91a}. The classical counterpart of
the modes \(es) is the following: From all directions particles with the impact
parameter $\rho_{\rm min}$ move towards the string. Due to the symmetry this
leads to a zero net radial flux but nonvanishing $\theta$-flux. The particle
density is then zero for $\rho <\rho_{\rm min}$ and $j^0_{cl} \approx N_0 (\nu
E_p /\pi p_{\perp}\rho)$ for large distances where $N_0$ is the total number of
particles. From this we can read off that the modes \(es) correspond to a
number of $N_0=1/4\pi^2E_p$ particles.

\medskip

The Maxwell equations in the cosmic string space-time with metric \(m), in the
Lorentz gauge, take a decoupled form
$$
\square_{\xi}{A_{\xi}} = 0\;, \quad {\rm with}\quad
\square_{\xi} = \Delta_{\xi} -\partial^{2}_{t t}\;,\eqno(maxdec)$$
$$
\Delta_{\xi} = {1 \over \rho} \partial_{\rho}( \rho \partial_{\rho})
- {1 \over \rho^2}L^{2}_3 + \partial^{2}_{z z}\;,\quad
L_3 = -i\partial_{\theta} + \xi \;\eqno(lap)$$
for independent spin-weighted components of the vector potential
\refto{Aliev89a},
$$
A_{\xi} = {1 \over \sqrt{2}}(A_{\rho} +{i\xi\over \rho}
A_{\theta})\quad {\rm if}\quad \xi=\pm 1\quad ;\quad
{\rm for}\; A_z, A_t \quad \xi=0\;.\eqno(axi)$$
In these terms the Maxwell field operator is
$$
A_{\xi}(t,x) = \int d\mu_j \left(f_{\xi}(j,x)c_{\xi} +
f^{\ast}_{-\xi}(j,x)c^{\dagger}_{-\xi}\right)\;\eqno(mo) $$
where the coefficients $c_{\xi}(j),\;c^{\dagger}_{\xi}(j)$ are annihilation
and creation operators for a photon with quantum numbers $j = (k_{\perp},
k_3, m, \xi)$ with commutation relations
$$
[c_{\xi}(j), c^{\dagger}_{\xi}(j^{\prime})] = \delta_{j, j^{ \prime}}
\;.\eqno(cr) $$
The photon normal modes are given by
$$
f_{\xi}(j,x) = {\sqrt \nu \over 2 \pi} \exp(i \nu m \theta + ik_3
z)J_{|\nu m + \xi|}(k_{\perp} \rho)
\exp\left(i{\pi \over 2}|m+ \xi|\right){1 \over
\sqrt{2\omega_{k}}}\exp(-i\omega_{k}t),\eqno(ms)$$
$\xi$ being a polarization state. They are the eigenfunctions for the set of
operators $\hat{k}_z,\; L_3=i\partial_{\theta}+\xi$ and are labeled by their
eigenvalues $k_3,\;l_3=\nu m+\xi.$ Notice that we add a phase factor too. The
modes \(ms) are normalized according to
$$
\int dx f^{\ast}_{\xi}(j,x)(i \buildrel \leftrightarrow \over
\partial_t) f_{\xi}(j^{\prime},x) = \delta_{j, j^{ \prime}}\;.\eqno(mn) $$

Now we discuss what the physical photon states are. To fix them we make use of
the Lorentz gauge condition which translates into
$$
\left({k_{\perp} \over \sqrt{2}}(c_{+} + c_{-}) + k_3 c_3-\omega_k c_0\right)
|\hbox{phys.state}>=0.  \eqno(lc1)$$
One can see that the operator
$$
c_l={k_{\perp}\over\omega_k }{c_{+}+c_{-}\over\sqrt{2}}+{k_3\over\omega_k}c_3
\;\eqno(lon)$$
corresponds to a polarization vector directed along $\vec{k}$. It is the
annihilation operator for longitudinal photons. With this operator the Lorentz
condition \(lc1) takes the simple form
$$
(c_l-c_0)|\hbox{phys.state}> = 0\eqno(lc2).$$
where $c_0$ is the annihilation operator for scalar photons. Annihilation
operators for transverse, physical photons can be defined up to a rotation
around $\vec{k}$. We fix them as follows
$$
c_{\sigma} = -i\;{(c_{+} - c{-}) \over \sqrt 2},\quad\quad
c_{\pi} = - {(c_{+} + c_{-}) \over \sqrt 2} {k_3 \over \omega_k} + c_3
{k_{\perp} \over \omega_k}. \; \eqno(trp)$$
Notice that operators $c_l,\;c_{\sigma}\;{\rm and}\;c_{\pi}$ obey the same
commutation relations \(cr).

\vskip2.5cm

\head{3. Cross section for bremsstrahlung process $e \rightarrow e + \gamma$}

\vskip0.25cm

\subhead{3.1. Matrix elements for the bremsstrahlung process}

\vskip0.25cm

In this section we evaluate the first order S-matrix element for the
bremsstrahlung of an electron moving freely in the flat but conical space-time
around the string. This was already done for a simplified model based on scalar
QED \refto{Audretsch91b}. Then we present the general expression for the
partial cross section for the bremsstrahlung process.

The QED interaction Lagrangian is
$$
L_{int} = -e\sqrt{-g}\bar \psi(x)A_{\mu}(x)\gamma^{\mu}(x)\psi(x) \eqno(qed)$$
where
$$
A_{\mu}(x) \gamma^{\mu}(x) =
\sqrt{2}\left[A_{+} \exp(i\nu \theta) \gamma^{+} + A_{-}
\exp(-i\nu \theta) \gamma^{-}\right] + A_{z}(x) \gamma^3\;\eqno(ag)$$
with
$$\eqalign{
&{\gamma^{\pm} = {1 \over 2}( \gamma^1 \mp i\gamma^2) = \pmatrix{
              0    & \sigma_{\pm} \cr
    - \sigma_{\pm} &       0 \cr}, \;
\gamma^3 = \pmatrix{
         0    & \sigma_3 \cr
    - \sigma_3 &       0 \cr}, \;}\cr
&{\sigma_{+} = \pmatrix{ 0 & 0 \cr
                        1 & 0 \cr}, \;
\sigma_{-} = \pmatrix{ 0 & 1 \cr
                        0 & 0 \cr}\;
\sigma_3 = \pmatrix{ 0 & 1 \cr
                        0 & -1 \cr}\;}\cr.}\eqno(gampm)$$
Let an ingoing electron with quantum numbers $j_p=(p_{\perp},p_3,l,s)$ emit a
photon with quantum numbers $j_k=(k_{\perp},k_3,m,\lambda)$ and an outgoing
electron take quantum numbers $j_q=(q_{\perp}, q_3,n,r).$ The matrix element of
this bremsstrahlung process for physical, transverse photon states $\lambda
=\sigma,\;\pi$ can be calculated more easily in terms of matrix elements
$M_{\lambda}$ for photon states with the polarization $\lambda=\pm ,\;3$,
$$
M_{\lambda} = -i<j_q, j_k|S^{(1)}|j_p> = -e \int {d^4x}\bar\psi_{e}(j_{q},x)
[c_{\lambda},\;A_{\mu}\gamma^{\mu}] \psi_{e}(j_{p},x) \eqno(me1)$$
where
$$
[c_{\pm},\; A_{\mu}\gamma^{\mu}] = \sqrt{2}\gamma ^{\mp} \exp( \mp i \nu
\theta) f^{\ast}_{\pm}(j_{k},x), \quad
[c_{3},\; A_{\mu}\gamma^{\mu}] = \gamma ^3 f^{\ast}_{0}(j_{k},x).$$
After a simple integration on $t,z$ and $\theta$ it becomes
$$
M_{\lambda} = -e \sqrt{\nu}{{\sqrt{pq}} \over{4\sqrt{2\omega_{k} E_{q}E_{p}}}}
\exp(i\eta){\delta(E_p-E_q- \omega_{k}) \delta(p_3 - q_3 - k_3)
\delta_{l,\;m + n}}\; m_{\lambda} \eqno(me2)$$
where $\eta=i(\pi /2)(|l|-|n|-|m|)$ and
$$\eqalign{
m_{+} =& \epsilon_{l}\epsilon_{m} sr\sqrt{2(p- sp_3)(q+rq_3)} R J_{-+},\;\cr
m_{-} =& \epsilon_{n}\epsilon_{m}\sqrt{2(p+sp_3)(q-rq_3)} R J_{+-},\;\cr
m_{3} =& \left[\sqrt{(p+sp_3)(q+rq_3)} r J_{-0} - \epsilon_{l}\epsilon_{n}
\sqrt{(p-sp_3)(q-rq_3)} s J_{+0}\right] R\;\cr}$$
with
$$
R={1\over\sqrt{E_{p}-M}\sqrt{E_{q}+M}}+{sr\over\sqrt{E_{q}-M}\sqrt{E_{p}+M}}$$
Here we denote
$$
J(\alpha, \beta) = \int_{0}^{ \infty}{d\rho}\rho J_{|\alpha|}(q_{\perp}\rho)
J_{|\beta|}(k_{\perp}\rho)
J_{|\alpha+\beta|}(p_{\perp}\rho)\eqno(jint)$$
and
$$
J_{+-}=J(\alpha_+,\beta_-),\quad J_{-+}=J(\alpha_-,\beta_+),\quad
J_{+0}=J(\alpha_+,\beta_0),\quad J_{-0}=J(\alpha_-,\beta_0)$$
where
$$\alpha_{\pm}=\nu (n+{1 \over 2})\pm {1 \over 2}, \quad
\beta_{\pm}=\nu m \pm 1, \quad \beta_0=\nu m.$$
The ingoing electron can emit the bremsstrahlung radiation when $p_{\perp} >
q_{\perp}+k_{\perp}$. For this case we obtain \refto{Gradshteyn80}
$$\eqalign{
J(\alpha,\beta)=& {2\over\pi p_{\perp}^2\cos(A+B)\cos(A-B)}\;
d(\alpha,\beta),\cr
d(\alpha,\beta)=& \Theta(-\alpha \beta)\sin[\pi
\hbox{min}(|\alpha|,|\beta|)]{\left({\sin{A}\over\cos{B}}\right)}^{|\alpha|}
{\left({\sin{B}\over\cos{A}}\right)}^{|\beta|}\cr}\eqno(jint2)$$
where
$$
q_{\perp} = p_{\perp}\sin{A}\cos{B}, \; k_{\perp} =
p_{\perp}\sin{B}\cos{A}\;.$$

We notice in \(me2) that energy as well as linear momentum along the
string direction are of course conserved.
The condition $l=n+m$ is the conservation law for the total angular
momentum projection along the string direction.

The matrix elements \(me2) contains the step function $\Theta(-\alpha\beta)
=\Theta(-n\cdot m)$ from \(jint2). This is also the case for scalar field
models \refto{Harari90,Audretsch91a,Audretsch91c} and for pair production in
conic QED \refto{Skarzh93}. The outgoing particles carry out total angular
momentum projections of opposite signs. This common feature for quantum
processes around cosmic strings can be explained in the framework of the
semiclassical picture presented in section 1. Opposite signs for angular
momentum projections means that virtual particles created from the vacuum move
along opposite sides of the string. In this case they can give their momentum
excess to the string and become real. The process is concentrated near the
string, and this assumes a localization mechanism for quantum processes in the
neighborhood of the string core, as it was discussed in
\refto{Audretsch91a,Audretsch91b}. Also, one can easily see that the matrix
element \(me2) is zero for the photon state with $m=0$.

 From the matrix element \(me2) we evaluate the partial cross section
per unit length of string for the bremsstrahlung
process for the physical states $\sigma,\; \pi$:
$$
d\sigma^{\lambda}_l=W_{\lambda}q_{\perp}dq_{\perp}k_{\perp}dk_{\perp}dq_3dk_3
\eqno(dp)$$
where
$$
W_{\lambda} = {{\nu e^2 pq}\over{32(2\pi)^{2}\omega_{k}
E_{q}E_{p}}}{\delta(E_p-E_q-\omega_{k})\;\delta(p_3-q_3-k_3)\;
\delta_{l,\;n+m}}\; |m_{\lambda}|^2\;,\eqno(W1)$$
and $\lambda=\sigma,\;\pi,$
$$
m_{\sigma}=-{i\over \sqrt{2}}(m_+ - m_-),\quad
m_{\pi}=-{m_+ + m_-\over\sqrt{2}}{k_3\over\omega_k}
+m_3{k_{\perp}\over\omega_k}.$$

The expressions \(dp) and \(W1) describe the distributions of intensity for
outgoing electron and photon over their quantum numbers, transverse and
longitudinal momenta, angular momentum projections and polarizations.

\vskip0.25cm

\subhead{3.2. Partial cross sections for the bremsstrahlung process}

\vskip0.25cm

In particular, we are interested in the partial cross sections and radiated
energy for the process in study. To evaluate them, we need to integrate over
final states. Now we will assume that the ingoing electron moves perpendicular
to the string, and we put $p_3=0$ to facilitate calculations. Due to the
invariance of the cosmic string metric \(m) under boost transformation along
the string direction we can easily recover the general case.

Summing over polarizations for the outgoing electron and averaging over them
for the ingoing electron yields
$$
{1\over 2}\sum_{s,r}|m_{\sigma}|^2 = {2 \over pq}\left(
(p_{\perp}^2+q_{\perp}^2-
k_{\perp}^2)(J_{-+}^2+J_{+-}^2)-4\epsilon_{l}\epsilon_{n}p_{\perp}q_{\perp}
J_{-+}J_{+-}\right), \eqno(sep1)$$
$$\eqalign{
{1\over 2}\sum_{s,r}|m_{\pi}|^2 =&{2 \over pq}\{{k_3^2 \over\omega_k^2}
[(p_{\perp}^2+q_{\perp}^2-k_{\perp}^2)(J_{-+}^2+J_{+-}^2)\cr
&+4\epsilon_{l}\epsilon_{n}p_{\perp}q_{\perp}J_{-+}J_{+-}+
4\epsilon_{l}\epsilon_{m}p_{\perp}k_{\perp}(J_{-+}J_{-0}+J_{+-}J_{+0})]\cr
&+{k_{\perp}^2 \over \omega_k^2}
[(p_{\perp}^2+q_{\perp}^2-k_{\perp}^2)(J_{+0}^2+J_{-0}^2)-
4\epsilon_{l}\epsilon_{n}p_{\perp}q_{\perp}J_{+0}J_{-0}]\}.\cr}\eqno(sep2)$$

With the $n,m$-dependent part $d(\alpha, \beta)$ of integral \(jint2) we
calculate the sums over angular momentum quantum numbers of outgoing particles,
and we find (see Appendix):
$$\eqalign{
&\sum_{n,m}\delta_{l,n+m}(d_{-+}^2+d_{+-}^2)=\;{1+ab^2 \over b\sqrt{a}}
F_{\nu}(a,b),\quad
\sum_{n,m}\delta_{l,n+m}(d_{-0}^2+d_{+0}^2)=\;{1+a \over \sqrt{a}}
F_{\nu}(a,b),\cr
&\sum_{n,m}\delta_{l,n+m}\epsilon_l \epsilon_n (d_{-+}^2+d_{+-}^2)=
\sum_{n,m}\delta_{l,n+m}\epsilon_l \epsilon_n (d_{-0}^2+d_{+0}^2)=\;
F_{\nu}(a,b),\cr
&\sum_{n,m}\delta_{l,n+m}\epsilon_l \epsilon_m (d_{-+}d_{-0}+d_{+-}d_{+0})=\;
{1+ab\over \sqrt{ab}}F_{\nu}(a,b)\cr}\eqno(snm)$$
where
$$\eqalign{
&F_{\nu}(a,b)=\Sigma_1 a^{\nu|l+{1\over2}|} + (ab)^{-{\nu\over2}}\Sigma_2
b^{\nu|l+{1\over2}|},\cr
&\Sigma_1 =\sum_{m=1}^{\infty} \sin^{2}(\pi \nu m)\;(ab)^{\nu m} =
{c(1+c)\sin^{2}{\pi \nu} \over (1-c)[(1-c)^2+4c\sin^{2}{\pi \nu}]},\cr
&\Sigma_2 =\sum_{m=1}^{\infty} \cos^{2}{\pi \nu (m-{1\over2})}\;(ab)^{\nu m} =
{{c[(1-c)^2\cos^2{\pi \nu \over2}+2c\sin^2{\pi \nu}]}
\over{(1-c)[(1-c)^2+4c\sin^2{\pi \nu}]}}\cr}$$
and
$$
a={\sin^2A \over \cos^2B}, \quad b={\sin^2B \over \cos^2A},\quad
c=(ab)^{\nu}.$$
Inserting this in \(W1) we get
$$
{1\over 2}\sum_{s,r,n,m}W_{\lambda} =
{\nu\;e^2\;\delta(E_p-E_q-\omega_{k})\;\delta(p_3-q_3-k_3) \over
16\pi^{4}\omega_{k}E_{q}E_{p}\;[p_{\perp}^4-2p_{\perp}^2(q_{\perp}^2+k_{\perp}^2
) +(q_{\perp}^2-k_{\perp}^2)]}\; w_{\lambda} F_{\nu}(a,b);,\eqno(Wls)$$
where
$$\eqalign{
w_{\sigma}=&(p_{\perp}^2+q_{\perp}^2-k_{\perp}^2){1+ab^2\over
b\sqrt{a}}-4p_{\perp}q_{\perp},\cr
w_{\pi}=&(p_{\perp}^2+q_{\perp}^2-k_{\perp}^2){1+ab^2\over b\sqrt{a}}+
4p_{\perp}q_{\perp}+4p_{\perp}k_{\perp}{1+ab\over \sqrt{ab}}\cr
&+{k_{\perp}^2 \over \omega_k^2}[(p_{\perp}^2+q_{\perp}^2-k_{\perp}^2){1+a
\over \sqrt{a}}-(p_{\perp}^2+q_{\perp}^2-k_{\perp}^2){1+ab^2\over
b\sqrt{a}}-8p_{\perp}q_{\perp}-4p_{\perp}k_{\perp}{1+ab\over
\sqrt{ab}}].\cr}\eqno(wls)$$

Next we need to integrate on $dq_{3}dk_{3}.$ One can easily calculate for
$p_3=0$
$$
I(f)=\int_{-\infty}^{\infty}dq_3\int_{-\infty}^{\infty}dk_3{f(\omega_k)\over
E_q\omega_k}\delta(E_p -E_q-\omega_k)\delta(p_3-q_3-k_3)=
I_0 f(p_{\perp}\omega)\eqno(intpqf)$$
where
$$
I_0={4\Theta(s)\over p_{\perp}^2\sqrt{s}},\quad
s=1-2{q^{2}_{\perp}+k^{2}_{\perp}\over
p_{\perp}^2}+{(q^{2}_{\perp}-k^{2}_{\perp})^2\over p_{\perp}^4}-4\epsilon
{k^{2}_{\perp}\over p^{2}_{\perp}}\eqno(intp0)$$
and
$$
\omega={p_{\perp}^2-q_{\perp}^2+k_{\perp}^2\over 2p_{\perp}
\sqrt{p_{\perp}^2+M^2}},\quad \epsilon ={M^2\over p^{2}_{\perp}}.$$

The final step would be the integration on $q_{\perp}$ and $k_{\perp}$, which
can not be done analytically for arbitrary ingoing electron energy. Thus we now
introduce a convenient parametrization that facilitates analytic approximations
in different energy regimes.

The partial cross section is
$$
\sigma_l^{\lambda}= {\nu\;e^2\over 4\pi^4\;E_{p}}\int q_{\perp}dq_{\perp}\int
k_{\perp}dk_{\perp}{\Theta(s)\; A^{\lambda}(p_{\perp},q_{\perp}) \over
[p^{4}_{\perp}-2p^{2}_{\perp}(q^{2}_{\perp}+k^{2}_{\perp}) +
(q^{2}_{\perp}-k^{2}_{\perp})^{2}]\;\sqrt{s}}\; F_{\nu}(a,b)\eqno(pcs)$$
where
$$\eqalign{
p_{\perp}^2 A^{\sigma}(p_{\perp},q_{\perp})=
&(p_{\perp}^2+ q_{\perp}^2-k_{\perp}^2){1+ab^2\over
b\sqrt{a}}-4p_{\perp}q_{\perp},\cr
p_{\perp}^2 A^{\pi}(p_{\perp},q_{\perp})=
&(p_{\perp}^2+q_{\perp}^2-k_{\perp}^2) {1+ab^2\over
b\sqrt{a}}+4p_{\perp}q_{\perp}+4p_{\perp}k_{\perp}{1+ab\over \sqrt{ab}}\cr
&+{4k_{\perp}^2(p_{\perp}^2+M^2)\over
(p_{\perp}^2-q_{\perp}^2+k_{\perp}^2)^2}\left[(p_{\perp}^2+q_{\perp}^2-k_{\perp}
^2){1+a \over \sqrt{a}}-(p_{\perp}^2+q_{\perp}^2-k_{\perp}^2){1+ab^2\over
b\sqrt{a}}\right.\cr
&-\left.8p_{\perp}q_{\perp}-4p_{\perp}k_{\perp}{1+ab\over
\sqrt{ab}}\right].\cr}\eqno(pcsS)$$
The convenient variables to the problem are $\omega$ and $x$,
$$
\omega={p_{\perp}^2-q_{\perp}^2+k_{\perp}^2\over 2p_{\perp}
\sqrt{p_{\perp}^2+M^2}},\quad
x={2k_{\perp}\sqrt{p_{\perp}^2+M^2}\over p_{\perp}^2-q_{\perp}^2+k_{\perp}^2}.
\eqno(ox)$$
Here $p_{\perp}\omega$ is the photon energy $\omega_k$ and $x$ is $\sin
\theta_k$ where $\theta_k$ is the angle between the photon momentum vector and
the string direction. In these variables we write the final closed expression
for the partial cross section
$$
\sigma_l^{\lambda}={\nu e^2\over 32\pi^4E_p}\cos^2{\pi\nu\over2}\int_0^1
dx\int_0^{\omega_{\rm max}} {d\omega}{x\over \omega (1-x^2v^2)\sqrt{1-x^2}} \;
B^{\lambda}(x,\omega)f_{\nu}(a,b) \eqno(pcsxo)$$
where $\omega_{\rm max}=v\;(1+\sqrt{1-x^2v^2})^{-1},\;
v={p_{\perp}\over\sqrt{p_{\perp}^2+M^2}}\;$ is the velocity of the ingoing
electron and
$$\eqalign{
B^{\sigma}(x,\omega)=&
{4(1-x^2v^2)\over x^2\sqrt{1-2{\omega\over v} +\omega^2x^2}}
\left[2(1-{\omega\over v})+\omega^2 x^2\right],\cr
B^{\pi}(x,\omega)=&
 {4(1-x^2v^2)\over x^2\sqrt{1-2{\omega\over v}+\omega^2x^2}}
\left[2(1-{\omega\over v})(1-x^2)+\omega^2 x^2\right]\cr
& +8(1-x^2){1-{\omega\over v} \over \sqrt{1-2{\omega\over v}+\omega^2x^2}},\cr
f_{\nu}(a,b)=&
 {c\over(1-c)[(1-c)^2+4c\sin^{2}{\pi\nu}]}\left\{
4(1+c)\;a^{\nu|l+{1\over2}|}\sin^{2}{\pi\nu \over 2} \right. \cr
&+\left.{1 \over \sqrt{c}}\;b^{\nu|l+{1\over2}|}
\left[(1-c)^2+8c\sin^{2}{\pi\nu \over 2}\right] \right\} \cr}\eqno(Af)$$
with
$$c=(ab)^{\nu}, \quad a={v-\omega(1+\sqrt{1-x^2v^2})
\over v-\omega(1-\sqrt{1-x^2v^2})},\quad
b={1-\sqrt{1-x^2v^2}\over 1+\sqrt{1-x^2v^2}}\;.
\eqno(abcxo)$$
The expression \(pcsxo) describes the energy and angular distributions for the
intensity of the bremsstrahlung radiation. It allows us to analyze correlations
between the energy and the direction of radiation.  We extract the factor
$\cos^2{\pi\nu\over 2}$ from $F_{\nu}(a,b)$ to stress that the partial cross
section vanishes, as it should be, for $\nu=1$, when there is no deficit angle.
But it is different from the analogous factor $\sin^2{\pi\nu}$ that
appears in the case of scalar particles. This difference may be understood as
the influence of the spin connection on the Dirac equation in curved metrics.

We now analyze the expression \(pcsxo) under different approximations. We will
always consider the realistic case of GUT cosmic strings
$$
\nu \approx 1+ \delta, \quad {\rm with}\quad \delta=4G\mu \ll 1\;.\eqno(gut)$$
Since $\delta$ is of order of the mass per unit length in Planck units,
it is reasonable to assume it to be small (for instance it is of order
$10^{-6}$ for GUT cosmic strings).

Under this approximation the partial cross section becomes
$$
\sigma_l^{\lambda}={e^2\delta^2\over 128\pi^2E_p}\int_0^1 dx\int_0^{\omega_{\rm
max}} {d\omega}{x\over \omega (1-x^2v^2)\sqrt{1-x^2}}\;
B^{\lambda}(x,\omega)f_{\nu}(a,b)\eqno(gutpcs)$$
with
$$\eqalign{
f_1(a,b)=&
{ab\over(1-ab)[(1-ab)^2+4ab\pi^2\delta^2]}\cdot \cr
&\left\{ 4(1+ab)\;a^{|l+{1\over2}|} + {1 \over
\sqrt{ab}}\;b^{|l+{1\over2}|}\left[(1-ab)^2+8ab\right]\right\}.\cr}\eqno(gutf)$$

\vskip0.25cm

\subhead{3.3. Approximations at different energy regimes}

\vskip0.25cm

Now we discuss the behavior of the partial cross section \(gutpcs) at low, high
and ultrahigh energies. The low energy case means $v \rightarrow 0.$  At high
energies there exist two different regimes. For one of them, when $1 \ll \gamma
\ll 1/\delta$ where $\gamma$ is the Lorentz factor $\gamma= (1-v^2)^{-1/2}$, we
can drop the term with $\delta^2$ in denominator \(gutf). On the contrary, at
ultrahigh energies, $\gamma \gg 1/\delta$, this term becomes dominant.

We first consider the low energy case. At $ v \rightarrow 0$ the interval of
integration over $\omega$ is very small since $\omega_{\rm max} \approx {v/2}
\ll 1.$ Then we have
$$\eqalign{
a & \approx (1-2{\omega\over v}),\quad b \approx {x^2v^2\over 4}\ll 1, \quad ab
\ll 1,\cr
B^{\sigma}& \approx {8(1-{\omega\over v})\over\sqrt{1-2{\omega\over v}}}\;
{1\over x^2},\quad
B^{\pi} \approx {8(1-{\omega\over v})\over\sqrt{1-2{\omega\over v}}}\;
{1-x^2\over x^2},\cr
&f_1(a,b) \approx (ab)\left(4a^{|l+{1\over 2}|}+{1\over \sqrt{ab}}b^{|l+{1\over
2}|}\right).\cr}\eqno(Afle)$$
One can see that the second term in $f_1(a,b)$ contributes at $l=0$ only.
At $l>0$ we find
$$\eqalign{
\sigma_l^{\lambda}=
&{e^2\delta^2v^2\over 16\pi^2M}\int_0^1dx {x
[\delta_{\lambda\sigma}+\delta_{\lambda\pi}(1-x^2)]\over(1-x^2v^2)
\sqrt{1-x^2}}\int_0^{\omega_{\rm max}}d\omega{(1-{\omega\over v})
(1-2{\omega\over v})^{l+1}\over\omega},\cr
\sigma_l^{\sigma}=
&3\sigma_l^{\pi}={e^2\delta^2v^2\over 32\pi^2M}
\left[\psi(1)-\psi(l+2)-\ln \eta_{\rm min}-{1\over 2(l+2)}\right],\cr
\sigma_l^{\sigma+\pi}=
&{e^2\delta^2v^2\over 24\pi^2M}
\left[\psi(1)-\psi(l+2)-\ln \eta_{\rm min}-{1\over 2(l+2)}\right]
\cr}\eqno(pcsle)$$
where $\psi(x)$ is the psi-function. We cut off the infrared divergent integral
at low frequency $\omega_{\rm min}= {v\eta_{\rm min}/2}$. An infrared
singularity of the same type arises in the first radiative correction to
S-matrix elements for scattering of an electron  by the cosmic string. These
singular terms cancel each other in the electron scattering cross section with
the emission of a number of soft photons. We will not discuss this interesting
problem now, and we plan to return to it for detailed investigation in future.
We are interested here in the energy behavior of the bremsstrahlung process
only. Note that the partial cross sections increase quadratically in velocity
independently on values of angular momentum projection $l$. This universal
energy behavior differs in this case from pair the production one and can be
naturally explained by the fact that there is no energy threshold for the
bremsstrahlung process.

At high energy, $\gamma \gg 1$, (and $v \approx 1$) both $a$ and $b\approx 1,
\;B^{\sigma},\;B^{\pi} \approx (1-x^2v^2)$ and
$$
f_1(a,b) \approx {16\over(1-ab)[(1-ab)^2+4\pi^2\delta^2]}\eqno(Afhe)$$
with
$$
(1-ab) \approx {2\sqrt{1-x^2v^2}\over 1-\omega}.$$
It means that the main contribution in the integrals arises from  values of $x
\approx 1$.

In this case one needs to be careful in finding the asymptotic behavior. We
dropped factors $a^l,\;b^l$ considering $l$ is finite. It is necessary to take
them into account at infinite $l$ if one is interested in the classical
consideration of bremsstrahlung process.

We still need to distinguish two different regimes at high energy: $1\ll
\gamma\ll 1/\delta$ and $\gamma\gg 1/\delta.$ The term between square brackets
in the denominator of \(Afhe) plays different roles wether $(1-ab)$ can be
smaller than $\delta$ or not. If the electron energy is not too high,
$\gamma\ll 1/\delta,$ we can neglect the term proportional $\delta^2$ and then
we find
$$
\sigma_l^{\lambda} \sim {e^2\delta^2\over M}\gamma. \eqno(pcshe)$$

Finally, at ultrahigh electron energy, $\gamma\gg 1/\delta$, the term
proportional $\delta^2$ dominates and then we obtain
$$
\sigma_l^{\lambda} \sim {e^2\over M}{1\over\gamma}\ln\gamma.\eqno(pcsuhe)$$

Notice that in both high energy regimes the cross section does not depend on
the electron quantum number $l$ that determines its angular momentum projection
along the string axis.This is actually only true up to some large value of $l$,
of order $p_{\perp}/M$, after which the approximations we were making break
down, and the cross section decreases with $l$. An heuristic, semiclassical
explanation for this property may arise from the following observation. The
classical analog of the cylindrical modes with given $l$ may be imagined as a
flux of particles incident upon the string from all directions with radius of
closest approach of order $r_{\rm  min}\approx l/p_{\perp}.$
\refto{Audretsch91a,Audretsch91b} This will be smaller than the Compton
wavelength $1/M$ of outgoing particles if $l<p_\perp/M$. In that case virtual
outgoing particles that move along opposite sides of the string, and thus are
able to extract momentum from it, will get hit by the electron. All values of
$l$ smaller than $p_\perp/M$ should then be equally efficient at
bremsstrahlung, while larger values should be less efficient. This seems to be
a common feature for all quantum processes around cosmic strings.

The obtained energy dependence for partial cross section should not continue
unbounded, and actually a cutoff is expected in a more realistic model, with
the conical singularity at the string core smoothed out. Also, we believe that
the perturbation theory will break down at ultrahigh energies when second
approximation can exceed the first one.

A striking feature is the independence of the cross section at ultrahigh
energies on $\delta$, it being a measure of the conical deficit angle. This
fact can also rouse a suspicion that the perturbation theory will break down at
ultrahigh energies. Again, an heuristic explanation may be found in the fact
that the string transfers momentum to ingoing particles of order $\delta
p_{\perp}$, and it is an effective parameter which determines the intensity of
quantum processes around the cosmic string \refto{Frolov88}. At ultrahigh
energy, when $p_{\perp} \gg M/\delta$, this effective parameter becomes
independent on the value of $\delta$.

\vskip0.5cm

\head{4. The radiated energy}

\vskip0.25cm

\subhead{4.1. The partial radiated energy}

\vskip0.25cm

It has been shown early \refto{Frolov88} that a charged particle moving
uniformly in the gravitational field of the cosmic string produces
electromagnetic radiation. In the framework of classical electrodynamics the
radiated energy reads \refto{Frolov88,Aliev89a}
$$\eqalign{
E_{\rm rad}=&-\int d^4x\sqrt{-g}j^{\nu}(x)A^{\rm rad}_{\nu;t}\cr
=&{e^2\delta^2\over 64\rho_{\rm min}}\gamma^2
\left[v(6v^2-1)+\gamma(4v^2+1){\rm arctg}(v\gamma)\right]\cr}\eqno(clre)$$
Quantum field theoretical calculation of bremsstrahlung process for scalar
electrodynamics was made in \refto{Audretsch91b}.

The averaged energy per unit length of the string carried off by a photon with
polarization $\lambda$ can be calculated by formula
$$
E^{\lambda}_{\rm rad} = \int \omega_k dw^{\lambda}.$$
Here the differential probability for the bremsstrahlung process for states
that correspond to 1 particle is given by $dw=4\pi^2 E_p d\sigma$. Thus we get
for the radiated energy of one electron moving in the cosmic string space-time
$$
E^{\lambda}_{\rm rad} = 4\pi^2E_p\int \omega_k d\sigma^{\lambda}. \;\eqno(re)$$
After integration over $dq_3dk_3$ the photon energy $\omega_k$ is replaced by
its effective value \(ox),
$$
p_{\perp}\omega={p_{\perp}^2-q_{\perp}^2+k_{\perp}^2\over 2
\sqrt{p_{\perp}^2+M^2}}.$$
So we have for the radiated energy
$$
E_{\rm rad}^{\lambda}= {\nu e^2 p_{\perp}\over
8\pi^2}\cos^2{\pi\nu\over2}\int_0^1 dx\int_0^{\omega_{\rm max}}
{d\omega}{x\over (1-x^2v^2)\sqrt{1-x^2}}
B^{\lambda}(x,\omega)f_{\nu}(a,b)\eqno(rade)$$
or, in GUT approximation,
$$
E_{\rm rad}^{\lambda}= {e^2 \delta^2 p_{\perp}\over 32}\int_0^1
dx\int_0^{\omega_{\rm max}} {d\omega}{x\over (1-x^2v^2)\sqrt{1-x^2}}
B^{\lambda}(x,\omega)f_1(a,b).\eqno(gutrade)$$

At low electron energies, $v \rightarrow 0$ one can easily obtain the
expression for the radiated energy from \(pcsle)
$$\eqalign{
&E^{\sigma}_{\rm rad}=3E^{\pi}_{rad}={e^2\delta^2 M\over 16} v^4 {2l+5 \over
(l+2)(l+3)},\cr
&E^{\sigma+\pi}_{\rm rad}={e^2\delta^2 M\over 12} v^4 {2l+5 \over
(l+2)(l+3)}\cr}\eqno(rele)$$
To compare with results of \refto{Frolov88,Aliev89a} for classical
electrodynamics we use the radius of closest approach \refto{Audretsch91a}
$\rho_{\rm min}=l/p_{\perp}$ and find from \(rele) at large $l$
$$\eqalign{
&E^{\sigma}_{\rm rad}=3E^{\pi}_{\rm rad}={e^2\delta^2v^3 \over 8}{1\over
\rho_{\rm min}},\cr
&E^{\sigma+\pi}_{\rm rad}={e^2\delta^2v^3 \over 6}{1\over \rho_{\rm
min}},\cr}\eqno(crele)$$

At not too high electron energy, $1 \ll \gamma \ll 1/\delta,$ the radiated
energy increases rapidly with electron energy,
$$
E_{\rm rad}^{\lambda} \sim e^2\delta^2 M \gamma^3, \eqno(rehe)$$
and then, at ultrahigh energy, $\gamma \gg 1/\delta,$ it goes asymptotically to
$$
E_{\rm rad}^{\lambda} \sim e^2 M \gamma \ln {\gamma}.\eqno(reuhe)$$
In both high energy regimes the radiated energy does not depend on the electron
angular momentum quantum number $l$. It is valid till not extremely large fixed
values $l$.

\vskip0.25cm

\subhead{4.2. Classical radiated energy}

\vskip0.25cm

In previous sections we have considered the bremsstrahlung process from quantum
electron in the eigenstate with given energy and fixed angular momentum quantum
number $j$. It is quite possible to derive the known expression \(clre) for
radiated energy from the general quantum formula \(rade). To do this it is
necessary to consider the situation when the Compton wave length of the
electron is much less than the radius of closest approach, $1/M \ll \rho_{\rm
min}=\nu |j|/p_{\perp},$ or $\nu |j| \gg v\gamma$. Under this approximation the
main contribution to the integral \(rade) will be determined by values of
variables with $a, b \approx 1$. It happens if $\omega \approx 0$ and in this
case we can neglect $b^{|j|}$. At small $\omega $ one can put $a \approx
1-2{\omega\over v} \sqrt{1-x^2v^2},$
$$\eqalign{
B^{\sigma} \approx {8(1-x^2v^2)\over x^2},\quad
B^{\pi} \approx {8(1-x^2)\over x^2},\cr
f_1(a,b) \approx {x^2 v^2 a^{|j|} \over \sqrt{1-x^2v^2}
[1-x^2v^2+\pi^2\delta^2x^2v^2]} \cr}$$
and
$$
a^{|j|} = \exp(|j|\ln a) \approx \exp\left(-|j|2{\omega\over v}
\sqrt{1-x^2v^2}\right)$$
at $j \rightarrow \infty.$ Then we obtain as final expression for the classical
radiated energy
$$
E_{\rm rad}^{\sigma+\pi} = {e^2 \delta^2 v^3\over 8\rho_{\rm min}} \int_0^1 dx
{x [2-x^2(1+v^2)]\over \sqrt{1-x^2} (1-x^2v^2)^2
[1-x^2v^2+\pi^2\delta^2x^2v^2]} \eqno(clrade)$$
which interpolates between different energy regimes. At not too high electron
energy, $1 \ll \gamma \ll 1/\delta $ we find the known result \(clre),
$$
E_{\rm rad} = {e^2 \delta^2  \over 64 \rho_{\rm min}}\gamma^2
\left[v(6v^2-1)+ \gamma (4v^2+1){\rm arctg}(v\gamma)\right]. \eqno(clre1)$$
At low energy we find from here
$$
E_{\rm rad} \approx {e^2 \delta^2 v^3 \over 6 \rho_{\rm min}}$$
and at high energy
$$
E_{\rm rad} \approx {5\pi e^2 \delta^2 \gamma^3\over 128 \rho_{\rm min}}.$$
Taking into account the term with $\delta^2$ we obtain the radiated energy at
ultrahigh electron energy
$$
E_{\rm rad} \approx {3e^2 \gamma \over 32\pi \rho_{\rm min}}.$$

\vskip0.5cm

\head{5. CONCLUSIONS}

\vskip0.25cm

We have shown that the lack of global linear momentum conservation in the plane
perpendicular to a cosmic string, which is a consequence of its conical
topology, permits bremsstrahlung from a single electron passing by the cosmic
string as well as others quantum processes although there is no local
gravitational field. Expressions \(gutpcs), \(pcsle), \(pcshe), \(pcsuhe) for
the cross sections of the bremsstrahlung process and \(gutrade), \(rele),
\(rehe), \(reuhe) and \(clrade) for emitted energy contain our quantitative
results for this process at different energy regimes and for alternative
quantum states of the ingoing electron and emitted photon.

Previous results of a similar nature were already obtained for a simplified
model based on scalar fields
\refto{Harari90,Harari91,Audretsch91a,Audretsch91b,Audretsch91c}. Their
extension to QED, though, turned out to be not so straightforward, both in its
technical details as well as in the energy dependence of the resulting cross
sections. The scalar model was based on a Lagrangian $\lambda\varphi\psi^2$,
with $\lambda$ being a coupling constant with the dimension of mass, $\varphi$
being a massless and $\psi$ a massive scalar field. The QED result corresponds
to the scalar case with $\lambda$ replaced by $e p_\perp$. In the high energy
regime this makes a significant difference.

The quantum processes for particles in the space-time of a cosmic string can be
regarded as a consequence of a kind of gravitational analog of the
Aharonov-Bohm effect. We mention that there is also a proper Aharonov-Bohm
interaction of fermions with the gauge potential around a cosmic string, in
models where the string carries non-integer fluxes
\refto{Alford89a,Gerbert89b,Perkins91a}.

It is very striking that the cross sections for all quantum processes for
scalar as well as for electromagnetic fields at ultrahigh energies,
eq.\(pcsuhe), are independent on the angular momentum quantum number $l$ and on
the cosmic string gravitational parameter $\delta$. This has two consequences:
Firstly, it means that the cross section of the process for ``plane wave''
electron states \refto{Harari90} will have quite different high energy behavior
than the partial cross sections have. This behavior should be cut off at
energies that probe the core of
the string, $p_\perp\approx \sqrt\mu$, where the metric is not well
approximated by that of eq.\(m), which has a conical singularity at the origin.
A real cosmic string has a smooth core, and its metric approaches a flat,
regular metric at the origin. The metric is that of a snub-nosed cone
\refto{Gregory87,Garfinkle85}. It was shown, for instance, that the $1/\rho$
self-force that a charged particle experiences in the conical space-time around
a string,\refto{Linet86,Smith90} is cut off at a distance of order the core
radius in the snub-nosed cone metric \refto{Perkins91b}. The falldown of the
thin-tube approximation was noticed in \refto{Kay91}. Secondly, the
disappearance of the gravitational parameter from expressions for cross
sections of processes allows to question about the validness of the
perturbation theory for ultrahigh energies. This makes the calculation of
second order approximations for quantum processes around cosmic strings
desirable .

\vskip0.5cm

\head{Appendix}

\vskip0.25cm

We calculate here the sums over angular momentum quantum numbers of outgoing
particles \(snm). Let us to bring the expression \(jint2) for $d(\alpha,\beta)$
to a more suitable form
$$\eqalign{
d(\alpha,\beta)=
&\Theta(-\alpha\beta)\sin[\pi\hbox{min}(|\alpha|,|\beta|)] x^{|\alpha|}
y^{|\beta|}\cr
=&\Theta(\alpha +\beta)
\left[-\Theta(-\beta)\sin(\pi\beta) x^{\alpha}y^{-\beta}-
\Theta(-\alpha)\sin(\pi\alpha) x^{-\alpha} y^{\beta}\right]\cr
&+\Theta(-\alpha -\beta)
\left[\Theta(\alpha)\sin(\pi\alpha) x^{\alpha}y^{-\beta}+
\Theta(\beta)\sin(\pi\beta) x^{-\alpha}y^{\beta}\right],\cr}$$
where we denote $x=\sqrt{a}=\sin A/\cos B,\;y=\sqrt{b}=\sin B/\cos A.$

For all possible combinations $\alpha_{\pm}$ and $\beta_{\pm}$ from \(jint) the
inequality $\alpha +\beta > 0$ means $l\geq 0$ and $\alpha +\beta < 0$ means
$l\leq 0$. We use $\Theta(\pm\beta)$ and $\Theta(\pm\alpha)$ to set limits for
$m$ and $n$, respectively.

Then we obtain
$$\eqalign{
d_{-+}=& d(\alpha_-,\beta_+)\cr
=&\Theta(l\geq 0)
\left[\Theta(m\leq -1)\sin\pi\nu m\;(xy)^{-\nu m}x^{{\nu (l+{1\over
2)}}}x^{-{1\over 2}}y^{-1}\right.\cr
&+\left.\Theta(n\leq -1)\cos\pi\nu (n+{1\over 2})\;(xy)^{-\nu n}(xy)^{-{\nu
\over 2}} y^{\nu (l+{1\over 2)}} x^{1\over 2} y\right]\cr
&+\Theta(l\leq -1)
\left[-\Theta(n\geq 0)\cos\pi\nu(n+{1\over 2})\;(xy)^{\nu n} (xy)^{{\nu \over
2}}y^{-\nu (l+{1\over 2})} x^{-{1\over 2}}y^{-1}\right. \cr
&-\left.\Theta(m\geq 1)\sin\pi\nu m\;(xy)^{\nu m} x^{-{\nu (l+{1\over 2)}}}
x^{1\over 2} y\right],\cr}$$
$$\eqalign{
d_{+-}=& d(\alpha_+,\beta_-)\cr
=&\Theta(l\geq 0)
\left[\Theta(m\leq -1)\sin\pi\nu m\;(xy)^{-\nu m}x^{{\nu (l+{1\over
2)}}}x^{{1\over 2}}y\right.\cr
&-\left.\Theta(n\leq -1)\cos\pi\nu (n+{1\over 2})\;(xy)^{-\nu n}(xy)^{-{\nu
\over 2}} y^{\nu (l+{1\over 2)}} x^{-{1\over 2}} y^{-1}\right]\cr
&+\Theta(l\leq -1)
\left[\Theta(n\geq 0)\cos\pi\nu(n+{1\over 2})\;(xy)^{\nu n} (xy)^{{\nu \over
2}}y^{-\nu (l+{1\over 2})} x^{{1\over 2}}y \right. \cr
&-\left.\Theta(m\geq 1)\sin\pi\nu m\;(xy)^{\nu m} x^{-{\nu (l+{1\over 2)}}}
x^{-{1\over 2}} y^{-1}\right]\cr}$$
$$\eqalign{
d_{-0}=& d(\alpha_-,\beta_0)\cr
=&\Theta(l\geq 0)
\left[-\Theta(m\leq -1)\sin\pi\nu m\;(xy)^{-\nu m}x^{{\nu (l+{1\over
2)}}}x^{-{1\over 2}}\right.\cr
&+\left.\Theta(n\leq -1)\cos\pi\nu (n+{1\over 2})\;(xy)^{-\nu n}(xy)^{-{\nu
\over 2}} y^{\nu (l+{1\over 2)}} x^{1\over 2}\right]\cr
&+\Theta(l\leq -1)
\left[-\Theta(n\geq 0)\cos\pi\nu(n+{1\over 2})\;(xy)^{\nu n} (xy)^{{\nu \over
2}}y^{-\nu (l+{1\over 2})} x^{-{1\over 2}}\right. \cr
&+\left.\Theta(m\geq 1)\sin\pi\nu m\;(xy)^{\nu m} x^{-{\nu (l+{1\over 2)}}}
x^{1\over 2} y\right],\cr}$$
$$\eqalign{
d_{+0}=& d(\alpha_+,\beta_0)\cr
=&\Theta(l\geq 0)
\left[-\Theta(m\leq -1)\sin\pi\nu m\;(xy)^{-\nu m}x^{{\nu (l+{1\over
2)}}}x^{{1\over 2}}\right.\cr
&-\left.\Theta(n\leq -1)\cos\pi\nu (n+{1\over 2})\;(xy)^{-\nu n}(xy)^{-{\nu
\over 2}} y^{\nu (l+{1\over 2)}} x^{-{1\over 2}}\right]\cr
&+\Theta(l\leq -1)
\left[\Theta(n\geq 0)\cos\pi\nu(n+{1\over 2})\;(xy)^{\nu n} (xy)^{{\nu \over
2}}y^{-\nu (l+{1\over 2})} x^{{1\over 2}}\right. \cr
&+\left.\Theta(m\geq 1)\sin\pi\nu m\;(xy)^{\nu m} x^{-{\nu (l+{1\over 2)}}}
x^{-{1\over 2}} y\right].\cr}$$
Now we can easily to compute the sums \(snm). We obtain, for example,
$$\eqalign{
&\sum_{n,m}\delta_{l,n+m}(d_{-+}^2+d_{+-}^2)= \cr
&\sum_{n,m}\delta_{l,n+m}
\left\{\Theta(l\geq 0)
\left[\Theta(m\leq -1)\sin^2\pi\nu m\;(ab)^{-\nu m}a^{{\nu (l+{1\over
2)}}}({1\over b\sqrt{a}}+b\sqrt{a})\right. \right.\cr
&+\left.\Theta(n\leq -1)\cos^2\pi\nu (n+{1\over 2})\;(ab)^{-\nu n}(ab)^{-{\nu
\over 2}} b^{\nu (l+{1\over 2)}}({1\over b\sqrt{a}}+b\sqrt{a})\right]\cr
&+\Theta(l\leq -1)
\left[\Theta(n\geq 0)\cos^2\pi\nu(n+{1\over 2})\;(ab)^{\nu n} (ab)^{{\nu \over
2}}b^{-\nu (l+{1\over 2})}({1\over b\sqrt{a}}+b\sqrt{a})\right. \cr
&+\left.\left.\Theta(m\geq 1)\sin^2\pi\nu m\;(ab)^{\nu m} a^{-{\nu (l+{1\over
2)}}}({1\over b\sqrt{a}}+b\sqrt{a})\right]\right\}.\cr}$$
The sums over $n$ and $m$ are computed independently and we obtain the first
line in \(snm). The other sums are calculated analogously.

\vskip0.5cm

\head{Acknowledgments}

\vskip0.25cm

V.S. thanks Prof.J.Audretsch and the members of the Relativity group at the
University of Konstanz for hospitality, collaboration and many fruitful
discussions.

This work was supported by the Deutsche Forschungsgemeinschaft.

\vskip0.5cm

\references
\oneandahalfspace

\refis{Aharonov59}Y. Aharonov and D. Bohm, \journal Phys. Rev.,
119,485,1959.\par
\refis{Bethe34}H. Bethe and W. Heitler, \journal Proc. Roy. Soc.,
A 146,83,1934.\par
\refis{DeWitt60}B. S. DeWitt and R. W. Brehme, \journal Annals of
Physics,9,220,1960.\par
\refis{Frolov88}V. P. Frolov, E. M. Serebryany and V. D. Skarzhinsky, in
   {\it Proceedings of the 4th Moscow Seminar on Quantum Gravity}, ed. by
   M. A. Markov, V. A. Berezin and V. P. Frolov, World Sci. (1988).\par
\refis{Serebryany89} E. M. Serebryany and V. D. Skarzhinsky,
``The electromagnetic radiation at the Aha\-ro\-nov\--Bohm scat\-te\-ring" in
{\it Pro\-cee\-dings of the Lebedev Physical Institute} {\bf 197} (1989)
181.\par
\refis{Harari90}D. D. Harari and V. D. Skarzhinsky, \plb 240,322,1990.\par
\refis{Harari91}D. D. Harari and V. D. Skarzhinsky, in {\it ``Proceedings of
the V Seminar on Quantum Gravity''}, Moscow 1990, ed. by M. A.
Markov, V. A. Berezin and V. P. Frolov, World Scientific (1991); and
in {\it ``Proceedings of
First International A. D. Sakharov Conference in Physics''}, Moscow
(1991).\par
\refis{Skarzh93}V. D.Skarzhinsky, D. D.Harari and U. Jasper, \prd 49,755,1994.
\par
\refis{Audretsch91a}J. Audretsch and A. Economou, \prd 44,980,1991. \par
\refis{Audretsch91b}J. Audretsch and A. Economou, \prd 44,3774,1991. \par
\refis{Audretsch91c}J.Audretsch, A. Economou and D. Tsoubelis, \prd
45,1103,1992.\par
\refis{Aliev89a}A. N. Aliev and D .V. Gal'tsov, \journal Annals of
Physics,193,142,1989.\par
\refis{Aliev89b}A. N. Aliev and D. V. Gal'tsov, {\it Zh.Eksp.Teor.Fiz.}, {\bf
96}, (1989), 3; [{\it Sov.Phys.JETP}, {\bf 69}, (1989), 1].  \par
\refis{Alford89a}M. G. Alford and F. Wilczeck, \prl 62,1071,1989.\par
\refis{Alford89b}M. G. Alford, J. March-Russell and F. Wilczeck,
\npb 328,140,1989.\par
\refis{Bezzera87}V. B. Bezzera \prd 35,2031,1987.\par
\refis{Vilenkin85}A. Vilenkin, \journal Phys. Rep., 121,263,1985.  \par
\refis{Gott85}J. Gott, \journal Astrophys. J., 288,422,1985.\par
\refis{Frolov87}V. P. Frolov and E. M. Serebryany, \prd 35,3779,1987. \par
\refis{Deser84}S. Deser, R. Jackiw and G. 't Hooft, \journal Annals of Physics,
152,220,1984. \par
\refis{Deser84}S. Deser, R. Jackiw and G. 't Hooft, \journal  Annals of
Physics, 152,220,1984. \par
\refis{Deser88}S. Deser and R. Jackiw, \journal Comm.Math.Phys., 118,495,1988 ;
G.'t Hooft, \journal  Comm. Math. Phys., 117,685,1988.\par
\refis{Dowker87}J. S. Dowker, \prd 36,3095,1987 ; J. S. Dowker, \prd
36,3742,1987.\par
\refis{Hiscock87}W. A. Hiscock, \plb 188,317,1990. \par
\refis{Henneaux84}M. Henneaux, \prd 29,2766,1984. ; S. Deser, \cqg
2,489,1985. \par
\refis{Gerbert89a}Ph. de Sousa Gerbert, \prd 40,1346,1989.\par
\refis{Kogan91}Y. I. Kogan and K. G. Selivanov, \journal Int.J.Mod.Phys.,
A6,59,1991.\par
\refis{Kaiser84}N. Kaiser and A. Stebbins, \journal Nature, 310,391,1984. \par
\refis{Perkins91a}W. B. Perkins, L. Perivolaropoulos, A. C. Davis, R. H.
Brandenberger and A. Matheson, \npb 353,237,1991.\par
\refis{Perkins91b}W .B. Perkins and A. C. Davis,
\npb 349,207,1991.\par
\refis{Kay91}B. S. Kay and U. M. Studer, \journal Commun. Math.
Phys.,139,103,1991.\par
\refis{Hagen90}C. R. Hagen, \prl 64,503,1990. and \prd
41,2015,1990. \par
\refis{Gradshteyn80}I. S. Gradshteyn and I. M. Ryzhik,{\it Table of integrals,
series and products}, Academic Press (1980).\par
\refis{Linet86}B. Linet, \prd 33,1833,1986. \par
\refis{Smith90}G. Smith, in {\it `` Formation and evolution of cosmic
strings''}, Proceedings of the Cambridge Workshop, ed. by G. W.
Gibbons, S. W. Hawking and T. Vachaspati, Cambridge Univ. Press (1990).\par
\refis{Vilenkin81}A. Vilenkin, \prd 23,852,1981.  \par
\refis{Ford81}L .Ford and A. Vilenkin, \journal J. Phys.
A,14,2353,1981.\par
\refis{Garfinkle85}D. Garfinkle, \prd 32,1323,1985.  \par
\refis{Gregory87}R. Gregory, \prl 59,740,1987.\par
\refis{Voropaev91}S. A. Voropaev, D. V. Gal'tsov and D. A. Spasov, \plb
267,91,1991.; \journal Europhys. Lett.,12,609,1990.\par
\refis{Helliwel86}T. M. Helliwel and D. A. Konkowski, \prd 34,1918,1987. \par
\refis{Gerbert89b}Ph. de Sousa Gerbert and R. Jackiw, \journal Comm. Math.
Phys., 124,229,1989. \par
\endreferences

\endpaper\end